\documentclass[
   ,final            
  ,numberedheadings 
]{aipproc}
\layoutstyle{6x9}
\newcommand{\be}{\begin{equation}}
\newcommand{\ee}{\end{equation}}
\newcommand{\bea}{\begin{eqnarray}}
\newcommand{\eea}{\end{eqnarray}}
\newcommand{\bean}{\begin{eqnarray*}}
\newcommand{\eean}{\end{eqnarray*}}
\newcommand{\non}{\nonumber}
\newcommand{\bc}{\begin{center}}
\newcommand{\ec}{\end{center}}
\newcommand{\bi}{\begin{itemize}}
\newcommand{\ei}{\end{itemize}}
\newcommand{\dis}{\displaystyle}
\newcommand{\ra}{\rightarrow}

\begin{document}

\title{QCD Prediction of $A_{TT}$ for Small $Q_T$ Dimuon Production
in $pp$ and $p\bar{p}$
Collisions}


\author{Hiroyuki Kawamura}{address=
{Radiation Laboratory, RIKEN, Wako 351-0198, Japan}}

\author{Jiro Kodaira$^{1,}$}
{address=
{Theory Division, KEK, Tsukuba 305-0801, Japan
\footnotetext[1]{Deceased on Sep.16, 2006.}}}

\author{Kazuhiro Tanaka}{address=
{Department of Physics, Juntendo University, Inba, Chiba 270-1695, Japan}}

\begin{abstract}
We present QCD prediction of double-spin asymmetries 
($A_{TT}$) in transversely polarized Drell-Yan process 
at small transverse momentum $Q_T$ of dimuon. 
Resummation of large logarithmic corrections, relevant in small $Q_T$
region, is performed up to next-to-leading logarithmic (NLL) accuracy.   
$A_{TT}$ at RHIC, J-PARC and GSI are studied numerically 
in the corresponding kinematic regions.
We show that the large $A_{TT}$ is obtained for small $Q_T$ 
and moderate energies.     
\end{abstract}

\maketitle
The transversity $\delta q(x)$ is one of the twist-2 distribution 
functions of nucleon, which represents
the distribution of transversly 
polarized quarks inside transversly polarized nucleon \cite{RS:79}. 
Although it is theoretically as important as other twist-2 distribution 
functions such as the unpolarized parton distributions ($q(x), g(x)$)  
and the helicity distributions ($\Delta q(x), \Delta g(x)$), 
very little has been known of it so far.
This is because $\delta q(x)$ is a chiral-odd function, and 
should be always accompanied 
with another chiral-odd function 
in physical observable, and therefore cannot be measured in inclusive DIS. 
Transversely polarized Drell-Yan (tDY) process, 
$p^{\uparrow}p^{\uparrow}
\ra \ell^+\ell^- X
$, $p^{\uparrow}\bar{p}^{\uparrow} \ra \ell^+\ell^- X$, 
is one of the processes where we can access the transversity 
by measuring the double-spin asymmetry: 
$A_{TT}\equiv\frac{\Delta_Td\sigma}{d\sigma} 
\equiv
\frac{d\sigma^{\uparrow\uparrow}-d\sigma^{\uparrow\downarrow}}
{d\sigma^{\uparrow\uparrow}+d\sigma^{\uparrow\downarrow}}$.
Since tDY is inclusive in the final state, it is in principle the cleanest 
process to access $\delta q(x)$.
However, at the 
RHIC-Spin experiments,  
the asymmetries are likely to be quite small \cite{MSSV:98}. 
This comes from the fact that the DY process at $pp$ collider probes 
the sea distributions which are likely to be small for the
transversity. 
Moreover, the rapid growth of unpolarized sea distributions 
enhances the denominator of 
$A_{TT}$ at low-$x$ which is typically probed by tDY at RHIC.
On the other hand, much larger asymmetries are expected at the proposed 
spin experiments at J-PARC and GSI, which are to be performed at lower
energies \cite{BCCGR:06,SSVY:05,DRAGO:06}. 

In this work, we study the asymmetries in $Q_T$ distribution of dimuon, 
especially at small $Q_T$ where the bulk of dimuon is produced.
At small $Q_T$, the cross section is not described correctly 
by the fixed order calculations since large logarithmic corrections 
such as $\alpha_s^n\log^m(Q^2/Q_T^2)/Q_T^2 ~(m\leq 2n-1)$ 
appear at each order of perturbation series.
These so-called ``recoil logs'' come from soft gluon emissions, 
and have to be resummed to all orders in $\alpha_s$ to make a reliable 
prediction of the cross section at small $Q_T$. 
The resummation is carried out in the impact parameter $b$ space, 
conjugate to $Q_T$ space, to enforce transverse-momentum conservation, 
and the resummed cross section is expressed as the Fourier transform 
back to the $Q_T$ space. 
Here we perform the resummation at the next-to leading logarithmic 
(NLL) accuracy, which corresponds to adding up the terms with 
$m=2n-1$ and $2n-2$, respectively, for all $n$.  At this level, 
the ``resummed part'' of the spin-dependent cross section 
of tDY, differential in invariant mass $Q$, transverse momentum 
$Q_T$ and rapidity $y$ of dimuon, and in the azimuthal angle 
$\phi$ of one of the outgoing leptons is given by \cite{KKST:05} 
($\sqrt{S}$ is the CM energy of hadron system),
\bea
\frac{\Delta_T d \sigma^{\rm NLL}}{d Q^2 d Q_T^2 d y d \phi}
=&& \!\!\!\!\!\!\!\!\!\!\!\!
\cos(2 \phi )
\frac{\alpha^2}{3\, N_c\, S\, Q^2}
\sum_{i}e_i^2 \int_0^{\infty} d b \frac{b}{2}
J_0 (b Q_T)e^{\, S (b , Q)}
\label{resum}
\\
\times&& \!\!\!\!\!\!\!\!\!\!\!\!\!
\Biggl[  ( C_{qq} \otimes \delta q_i )
           \left( x_1^0 , \frac{b_0^2}{b^2} \right)      
( C_{\bar{q} \bar{q}} \otimes \delta\bar{q}_i )
           \left( x_2^0 , \frac{b_0^2}{b^2} \right)
+ ( x_1^0 \leftrightarrow x_2^0 )\Biggr].
\non
\eea 
Here $J_0 (b Q_T)$ is a Bessel function, 
$b_0 = 2e^{-\gamma_E}$ with $\gamma_E$ the Euler constant.
The large logarithmic corrections are resummed into 
the Sudakov factor $e^{S (b , Q)}$ with 
$S(b,Q)=-\int^{Q^2}_{b_0^2/b^2}(d\kappa^2/\kappa^2)
\{\ln\frac{Q^2}{\kappa^2}A_q(\alpha_s(\kappa))+B_q(\alpha_s(\kappa))\}$.
Perturbatively calculable quantities, i.e.,  $A_q$, $B_q$ and 
the coefficient functions $C_{qq}(z)$ and $C_{\bar{q}\bar{q}}(z)$, 
are expressed as power series in $\alpha_s$, and the their explicit forms
necessary at the NLL accuracy are found in Ref.\cite{KKST:05}. 
$x_{1,2}^0 = (Q/\sqrt{S})e^{\pm y}$, and 
$\delta q_i(x,\mu^2)$ is the transversity of $i$-th
flavour quark at the $\overline{\rm MS}$ scale $\mu$. 
The singularity in $b$-integration, due to Landau pole 
in $\alpha_s(\kappa)$, is taken care of by ``contour deformation
method'' introduced in \cite{LKSV:01}. 
Correspondingly, the nonperturbative effects are included 
by the replacement $e^{S(b,Q)}\ra e^{S(b,Q)-g_{NP}b^2}$ in (\ref{resum})
\cite{CSS:85,LKSV:01,BCDeG:03}, 
with a non-perturbative parameter $g_{NP}$, which (roughly speaking) 
parameterizes the intrinsic $k_T$-distribution of quarks inside nucleon. 
Then we combine the resummed 
part (\ref{resum}), 
which embodies the logarithmically enhanced contributions for small $Q_T$
to all orders,  
with the ``residual part'' of the fixed-order cross section, 
which is not associated with such logarithmic
enhancement.
To perform this consitently, we need the leading order (LO) 
tDY cross section at finite $Q_T$;
this is of $O(\alpha_s )$ and is 
obtained as QCD prediction at large $Q_T$~\cite{KKST:05}.
The matching of the NLL formula (\ref{resum}) with 
the corresponding component in the LO cross section is 
performed at intermediate $Q_T$ following the formulation 
of Ref.\cite{BCDeG:03}, to ensure no double counting for all $Q_T$,
and we obtain the ``NLL+LO'' prediction of the tDY cross section, 
$\Delta_Td\sigma^{\rm NLL+LO}/(dQ^2dQ^2_Tdyd\phi)$, which has a uniform 
accuracy over the entire range of $Q_T$ \cite{KKST:05}.
The ``NLL+LO'' cross section of unpolarized Drell-Yan process,
$d\sigma^{\rm NLL+LO}/(dQ^2dQ^2_Tdyd\phi)$,  
is obtained in the same way, 
utilizing the results in the literature \cite{CSS:85,AEGM:84}. 

Taking the ratio of these cross sections, we obtain the double-spin 
asymmetries:
\bea
\dis
A_{TT}=\left[\Delta_Td\sigma^{\rm NLL+LO}/dQ^2dQ_T^2dyd\phi\right]
\left/\left[d\sigma^{\rm NLL+LO}/dQ^2dQ_T^2dyd\phi\right]\right..
\label{asym}
\eea
In order to calculate $A_{TT}$ numerically, we need to assume 
a model of the transversity $\delta q_i(x)$ for the numerator. 
Here we take a model used in \cite{MSSV:98}, which saturates 
the Soffer bound \cite{Soffer:95} as 
$\delta q_i(x,\mu^2_0)=[q_i(x,\mu^2_0)+\Delta q_i(x,\mu^2_0)]/2$ 
at the low input scale $\mu_0\sim 0.6$GeV and is evolved to 
the higher $\mu^2$ with NLO DGLAP kernel \cite{KMHKKV:97};
as the inputs, we use GRV98 \cite{GRV:98} for $q_i(x,\mu_0^2)$ 
and GRSV2000 \cite{GRSV:00} for $\Delta q_i(x,\mu^2_0)$.
Correspondingly, the GRV98 distributions
are used for calculating the denominator of (\ref{asym}). 
The non-perturbative parameter $g_{NP}$ are taken to be common 
in the numerator and the denominator of (\ref{asym}), and we use 
$g_{NP}=0.5$ GeV$^2$ which is consistent with the result of 
\cite{KS:03}.~\footnote[2]{
The $g_{NP}$-dependence of the polarized and 
unpolarized cross sections almost cancels in $A_{TT}$ of (\ref{asym}) 
in the range $g_{NP}=0.3$-0.8 GeV$^2$.} 
 
\begin{figure}
\hspace*{-0.3cm}
 \includegraphics[height=5.1cm]{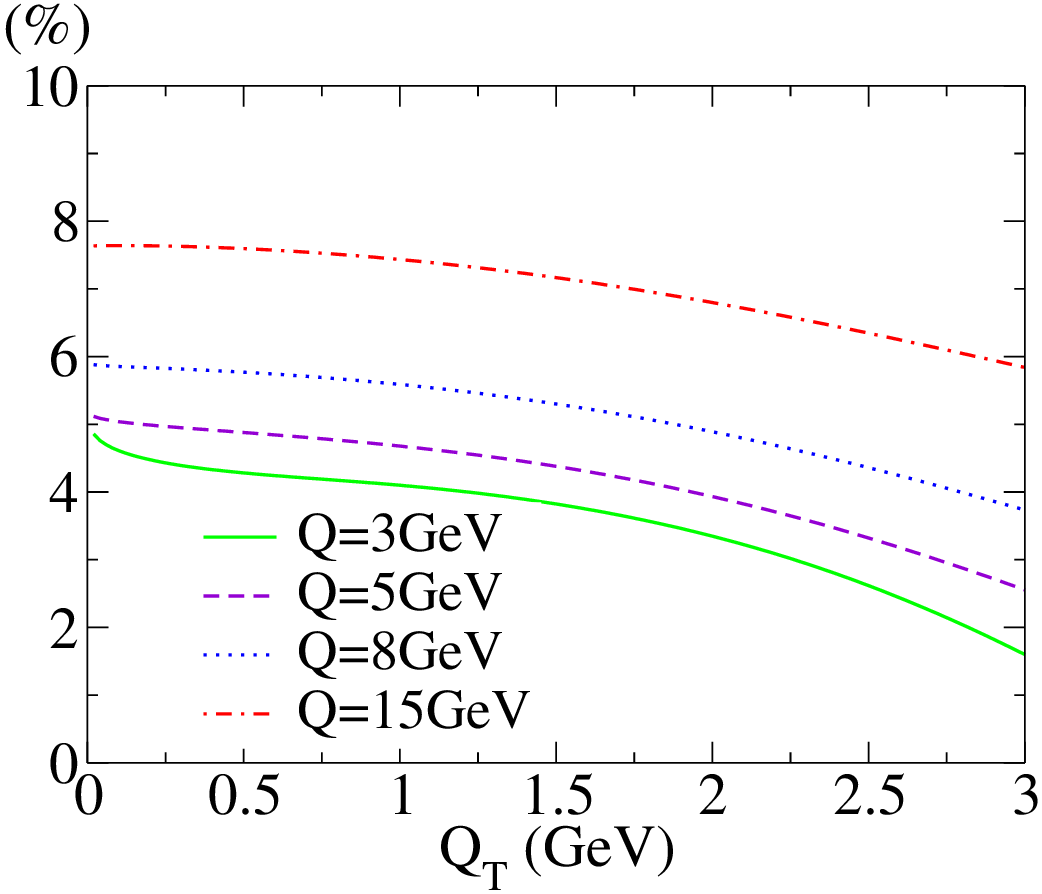}
\hspace{1.3cm}
 \includegraphics[height=5.1cm]{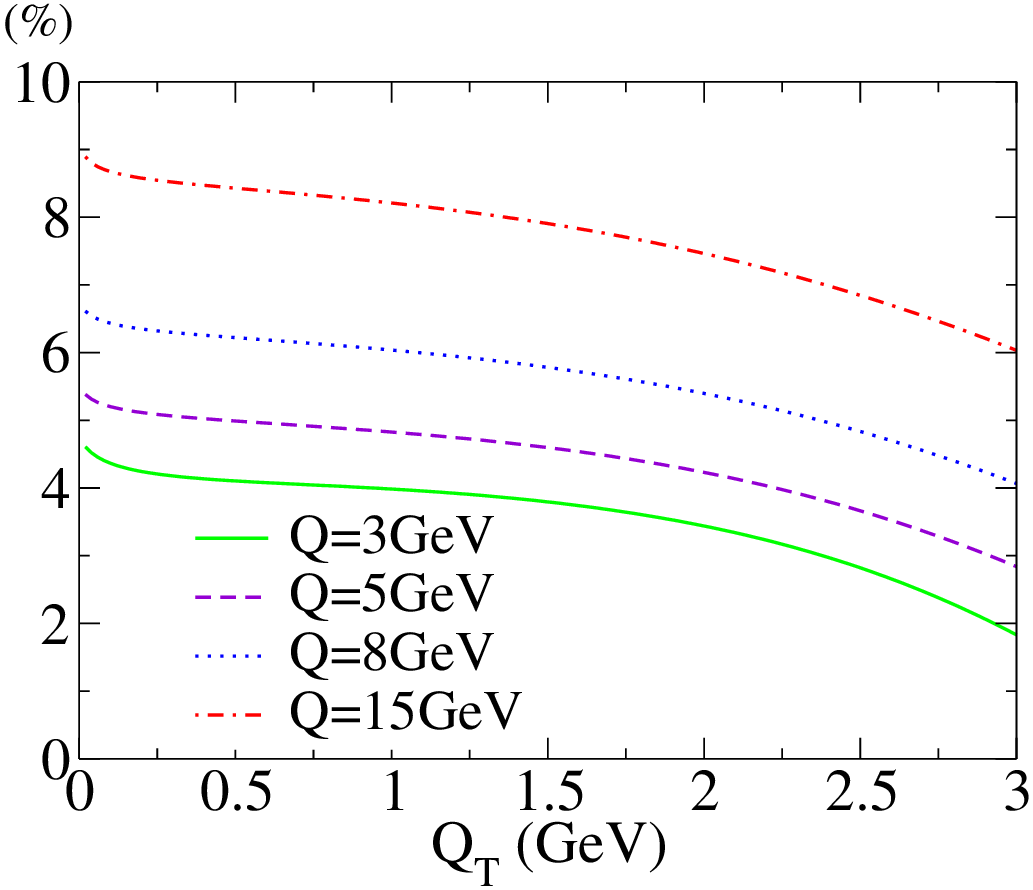}
 \caption{$A_{TT}$ at NLL+LO accuracy in $pp$ collision at RHIC kinematics:  
          $\sqrt{S}=200$ GeV, $Q=3,5,8,15$ GeV and $\phi=0$ 
          with $y=0$ (left panel) and $y=2$ (right panel).}
\end{figure}

\begin{figure}
\hspace*{-0.3cm}
 \includegraphics[height=5.1cm]{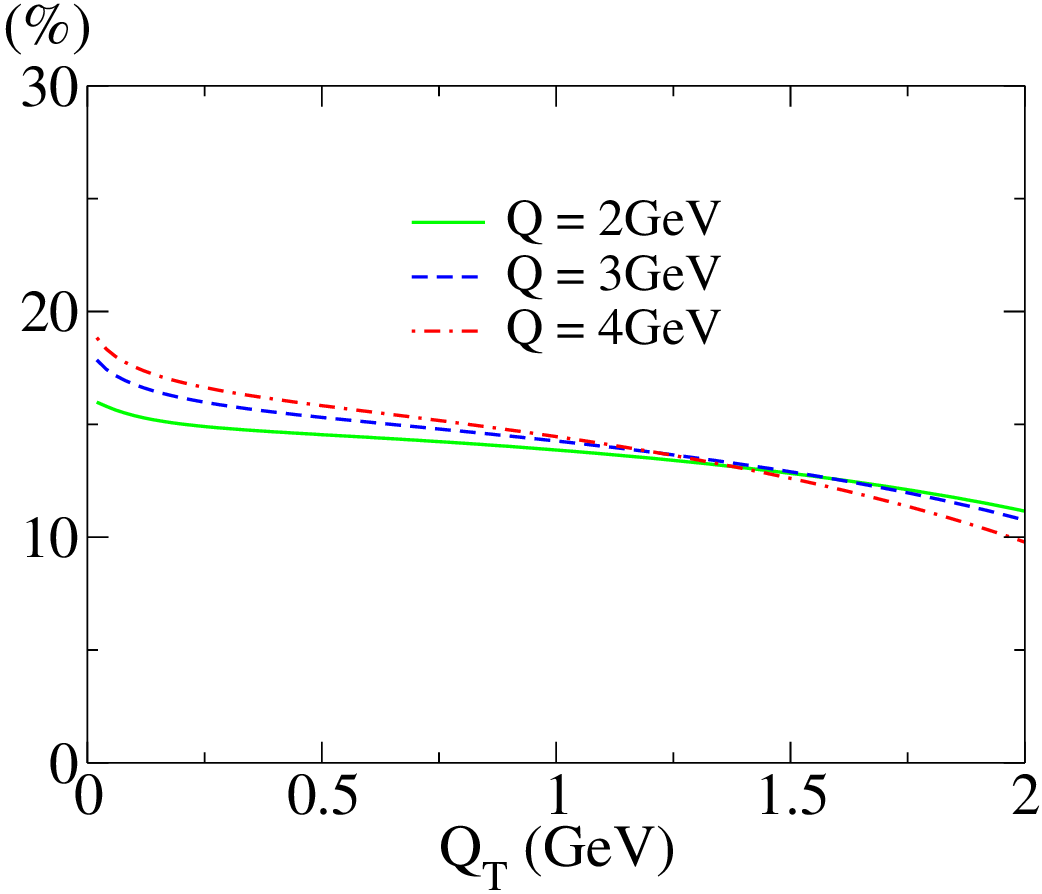}
\hspace{1.3cm}
 \includegraphics[height=5.1cm]{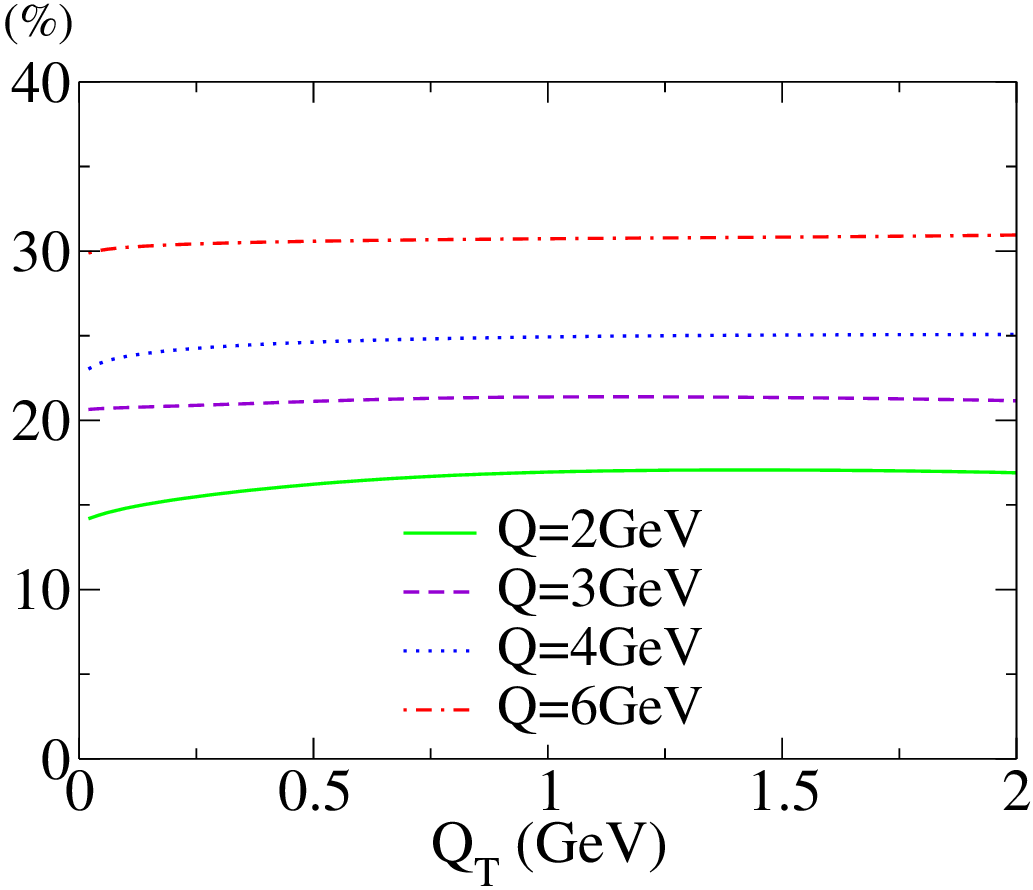}
 \caption{ $A_{TT}$ at NLL+LO accuracy. Left: 
          $\sqrt{S}=10$ GeV, $Q=2,3,4$ GeV, $\phi=0$, 
          and $y=0$ for $pp$ collision at J-PARC kinematics. Right: 
          $\sqrt{S}=14.5$ GeV, $Q=2,3,4,6$ GeV, $\phi=0$, and
          $y=0$ for $p\bar{p}$ collision at GSI kinematics.}
\end{figure}
 
Fig.~1 shows $A_{TT}$ in $pp$ collision at RHIC kinematics: 
$\sqrt{S}=200$ GeV, $Q=3,5,8,15$ GeV and $\phi=0$, with $y=0$ 
(left panel) and $y=2$ (right panel).
In this case, $A_{TT}$ are about 4-8\% 
and rather flat in the small $Q_T$ region where the resummed 
cross sections are dominant in both the numerator and denominator of (\ref{asym}). 
We have smaller $A_{TT}$ for smaller $Q$ due to the 
growth of the sea distributions in the denominator of (\ref{asym})
at small $x$.
We obtain slightly larger 
$A_{TT}$ for $y=2$ compared with the $y=0$ case, 
and it appears that generically the $y$-dependence of $A_{TT}$ is small. 
 
The left panel of Fig.~2 is same as Fig.~1, but at J-PARC kinematics: 
$\sqrt{S}=10$ GeV, $Q=2,3,4$ GeV, $\phi=0$ and $y=0$. 
In this kinematics, the parton distributions at medium $x$ 
($x_{1,2}^0=0.2$-$0.4$) are probed, so that 
the transversity distributions in the numerator of (\ref{asym})
are larger than the RHIC case while the growth of the sea distributions 
in the denominator is not significant.
Therefore we obtain much 
larger $A_{TT} \sim15$\% than the RHIC case of Fig.~1.
$A_{TT}$ are again flat as functions of $Q_T$,
and the dependence on $Q$ is also weak.

The right panel of Fig.~2 shows $A_{TT}$ in $p\bar{p}$ collision at GSI 
kinematics: $\sqrt{S}=14.5$ GeV, $Q=2,3,4,6$ GeV, $\phi=0$ 
and $y=0$, where $A_{TT}$ are dominated by valence distributions at medium $x$.
The largest $A_{TT}$ of 15-30\% are obtained in this case. 
The results are extremely flat as functions of $Q_T$.
Integrating the numerator and the denominator of (\ref{asym}) 
over $Q_T^2$, we reproduced the NLO asymmetries given by 
Barone et al.\cite{BCCGR:06}. 
We also calculated the $y$-dependence of the results in Fig.~2, and it turns out to
be small for both J-PARC and GSI cases.

To summarize,
we have calculated the double-spin asymmetries $A_{TT}$ for small $Q_T$
DY pair production in high-energy $pp$ collisions at RHIC
and in moderate-energy $pp$ and $p\bar{p}$ collisions at J-PARC and GSI.
Our results demonstrate that $A_{TT}$ reach a finite value at $Q_T = 0$
through the flat behavior in the small $Q_T$ region, reflecting 
that the soft gluon resummation to the NLL level has universal structure 
for the polarized and unpolarized DY \cite{KKST:05}, and also reveal that $A_{TT}$ are 
large enough to be experimentally measured, especially in moderate energies.



\begin{theacknowledgments}
We would like to thank Werner Vogelsang, Hiroshi Yokoya and 
Stefano Catani for valuable discussions and comments.
The work of J.K. and K.T. was supported by the Grant-in-Aid 
for Scientific Research Nos. C-16540255 and C-16540266.  
\end{theacknowledgments}

\end{document}